\documentclass[aps,prb,amsmath,amssymb,twocolumn]{revtex4}

\usepackage{graphicx,bm}

\begin{document}
\title{Quantum Hall plateau transition in the lowest Landau level of disordered graphene}
\author{Pallab Goswami}
\affiliation{Department of Physics and Astronomy\\ University of California Los Angeles, Los Angeles, California, 90095-1547}
\author{Xun Jia}
\affiliation{Department of Physics and Astronomy\\ University of
California Los Angeles, Los Angeles, California, 90095-1547}
\author{Sudip Chakravarty}
\affiliation{Department of Physics and Astronomy\\ University of California Los Angeles, Los Angeles, California, 90095-1547}
\date{\today}

\begin{abstract}
 We investigate, analytically and numerically, the effects of disorder on the density of states and on the localization properties of the relativistic two dimensional fermions
 in the lowest Landau level. Employing a supersymmetric technique, we calculate the exact  density of states for the
 Cauchy (Lorentzian)  distribution for various types of disorders.  We use a  numerical technique
to establish the localization-delocalization (LD) transition in the lowest Landau level. For some types of disorder the LD  transition is shown to belong to a different universality class, as compared to the corresponding nonrelativistic problem. The results are  relevant to the integer quantum Hall plateau transitions  observed in graphene.

 \end{abstract}
\pacs{PACS numbers:}

\maketitle
\section{Introduction}
Recent  experiments \cite{Novoselov1} have unravelled a fascinating set of phenomena in atomically thin layer of hexagonally arranged carbon atoms known as graphene.~\cite{review}
The quasiparticles
of graphene are $(2+1)$-dimensional massless Weyl fermions. \cite{Wallace,
Semenoff} In the context of condensed matter physics their properties are strikingly different from non-relativistic fermions. And phenomena that are hard to realize for the relativistic case, such as the Klein paradox or the {\em Zitterbewegung} are accessible in graphene.~\cite{Klein} It is perhaps not an exaggeration to remark that  many subtleties and a rich set of phenomenology are waiting to be discovered.

\subsection{Quantum Hall Effect}
A highlight has been the observation of an unconventional
quantum Hall effect  \cite{zhang1, novoselov2, zhang2} and the corresponding theoretical
development.\cite{gusynin1, gusynin2, peres, Brey, Abanin,
nomura, goerbig, alicea, gusynin3, herbut} In graphene  the filling fractions
are $\nu_{\mathrm{f}}=\pm 4(n+\frac{1}{2})$ for magnetic field $B<9T$, where $n$ is an integer.\cite{zhang1,
novoselov2, zhang2} The factor of $4$ comes from the two fold spin
degeneracy and the two fold nodal degeneracy of the Landau levels. The
Zeeman splitting is
negligible compared to the cyclotron frequency and the disorder
broadening of the Landau levels. The factor of half is due to  a zero mode in the Landau level spectrum of Dirac
fermions.\cite{gusynin1, gusynin2, peres}

For stronger
magnetic fields, $20T<B<45T$, plateaus appear at $\nu_{\mathrm{f}}=0,\pm1,
\pm2q$, where $q$ is an integer.\cite{zhang2}  The plateaus at
$\nu_{\mathrm{f}}=0,\pm1$ can be explained by the lifting of
both the spin and the nodal degeneracies in the lowest Landau level (LLL),
but those  at $\nu_{\mathrm{f}}=\pm4,
\pm6,...$ reflect only the removal of spin degeneracy in
higher Landau levels. The removal of nodal degeneracy requires  electron-electron interaction.
Mechanisms suggested
include $SU(4)$ ferromagnetism\cite{nomura, goerbig}, sublattice
symmetry breaking due to short range interactions\cite{alicea} and
the generation of a mass gap by magnetic catalysis. \cite{gusynin3,
herbut} $SU(4)$
quantum hall ferromagnetism predicts plateaus at all odd integer
filling fractions. However, apart from
$\nu_{\mathrm{f}}=\pm1$, the  plateaus at
$\nu_{\mathrm{f}}=\pm 3,\pm5,..$ have not yet been observed.

\subsection{Localization-delocalization transition}

The special quantization rules in graphene are explained by the
relativistic Landau levels, modified perhaps by interactions, but  for the existence of Hall
plateaus the Laughlin argument is
necessary.\cite{laughlin} According to this argument the
extended states at the center of a  Landau band are  separated by
the localized states elsewhere. If the Fermi energy falls in the
mobility gap, the plateaus are explained by a gauge invariance argument that is remarkably robust.  The
underlying phenomenon,  therefore, is a localization-delocalization (LD)
transition at the band center.\cite{laughlin, halperin, levine,
pruisken} The   conventional integer quantum Hall (IQH) plateau transition
has been  widely studied,  and it is known
 that the localization length exponent
$\nu\approx \frac{7}{3}$.\cite{aoki, huckestein1, huo, liu, chalker1,khmel,wei1,koch}  Can we prove that the
same argument applies to graphene, and, if so, does the LD
transition belong to the same universality class?

\subsection{Disorder and Dirac fermions}
 In the absence of a magnetic field,
 Dirac fermions in the presence of disorder  have been widely studied in
systems as varied as gapless semiconductors,~\cite{fradkin1}
gapless superconductors,~\cite{Nersesyan,  Zirnbauer} and IQH
plateau transitions.~\cite{fradkin2, Ludwig} As compared to  nonrelativistic fermions, the localization problem of
Dirac fermions is richer because of a
number of discrete symmetries. More
specifically, if the disorder is particle-hole symmetric,  for
example a random gauge field,  the LD transition takes place at
zero energy and is reflected in the single particle density of states (DOS),
 in contrast to the conventional metal-insulator
transition where  the DOS is smooth
through the LD transition. Surprisingly, there is a line of fixed
points with continuously varying exponents depending on
the disorder coupling constant. \cite{Ludwig, Nersesyan,
Zirnbauer} Some of the unusual behavior of disordered Dirac
fermions may be expected to realize in graphene. One such
effect that has received considerable attention
is the weak (anti)-localization phenomenon.\cite{Suzuura, McCann,
Nomura2, Morozov, Morpugo} However,  relativistic Landau
levels in the presence of disorder have not yet received much
attention.~\cite{peres, Ando, sheng,Haldane} Here we provide a reasonably
complete study of the possible effects.

There is another important reason why  LD
transitions in the relativistic Landau level should be carefully
analyzed. In the conventional IQH effect, the
spin-degenerate plateau transition corresponds to
$\nu\sim4.6$ when it is assumed that the LD transition takes place
at a single energy at the band center.\cite{wei2, engel, hwang,
wei3} This has led to intense theoretical investigation of the LD
transition in the spin-degenerate Landau band. \cite{chalker2,
chalker3, wang, hanna, kagalovsky} When spin-orbit scattering is
included, the LD transition is found to occur at two
distinct energies,  away from the band center. Scaling
analysis  about these distinct energies
provide, once again, that $\nu\approx 7/3$, as in the spin-polarized system. The
scaling about a single energy at the band center leads
to the effective exponent $\nu \sim 4.6$. One should anticipate a
similar discrepancy between the
spin and the nodal polarized IQH effect and the fourfold degenerate IQH effect in
graphene.

\subsection{Graphene in the lowest Landau level}

For simplicity we shall concentrate on the spin
polarized lowest Landau level (LLL) of graphene and analyze the LD transition both in
the presence and in the absence of nodal degeneracy.
An interesting example of  a controlled analytic calculation in
the disordered Landau level problem  is the
DOS in the LLL. This was first computed exactly by
Wegner~\cite{Wegner} by examining the Euler trails of the impurity
diagrams for the white noise disorder and was subsequently
extended by Brezin {\em et al.}~\cite{Brezin} by using a
supersymmetric (SUSY) technique. Here we also obtain some exact
results for the DOS in the disordered relativistic LLL  using SUSY
techniques.  

The most general model  of disorder consists of a
random potential, a random mass, a random gauge field and a random
internode scattering; however,  the random gauge field leaves the
LLL unperturbed. After projection to the spin-polarized LLL, we
study the following Hamiltonian:
\begin{equation}
{\hat{H}}_{\textrm{LLL}}=m\eta_{3}+\sum_{j=0}^{3}V_{j}(\vec{r})\eta_{j}  , \label{eq:LLL}
\end{equation}
where $V_{0}(\vec{r})$, $V_{3}(\vec{r})$ represent potential and mass
disorders respectively and $V_{1}(\vec{r})$ and $V_{2}(\vec{r})$
describe internode scattering effects. A mass $m$ of the
fermions have been included to study the effect of the removal of the
nodal degeneracy. For simplicity we have omitted the constant
Zeeman energy. The $2\times 2$ matrix $\eta_{0}$ is the identity matrix and
$\eta_{1}$, $\eta_{2}$ and $\eta_{3}$ are the three Pauli matrices.

\subsection{Summary of results}

Because of a large number of cases involved, it is useful to
summarize the results for the LD transition. Let $g_{0}$, $g_{3}$,
$g_{1}$ and $g_{2}$ denote the widths of the Gaussian random distributions corresponding to the random potential,  random mass, and random internode scatterings, respectively.

\subsubsection{$m=0$}
The list of possible cases are:
\begin{enumerate}
\item $g_{0}\neq 0$ and $g_{3}=g_{1}=g_{2}=0$.
\item  $g_{3}\neq 0$, $g_{0}=g_{1}=g_{2}=0$.
\item $g_{2}\neq 0$, $g_{0}=g_{3}=g_{1}=0$.
\item $g_{1}\neq 0$, $g_{0}=g_{3}=g_{2}=0$.
\item $g_{0}\neq 0$, $g_{3}\neq 0$ and $g_{1}=g_{2}=0$.
\item  $g_{0}\neq 0$, $g_{2}\neq 0$ and $g_{3}=g_{1}=0$.
\item  $g_{0}\neq 0$, $g_{1}\neq 0$ and $g_{3}=g_{2}=0$.
\item $g_{3}\neq 0$, $g_{2}\neq 0$, $g_{0}=g_{1}=0$.
\item $g_{3}\neq 0$, $g_{1}\neq 0$, $g_{0}=g_{2}= 0$.
\item  $g_{2}\neq 0$, $g_{1}\neq 0$ and $g_{0}=g_{3}=0$.
\item  $g_{0}\neq 0$, $g_{3}\neq 0$, $g_{2}\neq0$ and $g_{1}=0$.
\item  $g_{0}\neq 0$, $g_{3}\neq 0$, $g_{1}\neq0$ and $g_{2}=0$.
\item  $g_{0}\neq 0$, $g_{2}\neq 0$, $g_{1}\neq0$ and $g_{3}=0$.
\item  $g_{3}\neq 0$, $g_{2}\neq 0$, $g_{1}\neq0$ and $g_{0}=0$.
\item  $g_{0}\neq 0$, $g_{3}\neq 0$, $g_{2}\neq0$ and $g_{1}\neq0$.
\end{enumerate}

In the cases
(1) and (2), when disorder does not mix the two nodes, the LD
transitions belong to the conventional IQH universality class
with $\nu \approx 7/3$. It is interesting to note that mass
disorder produces LD transition in the LLL, whereas for zero magnetic  field random mass is known to be an
irrelevant perturbation for the $(2+1)$-dimensional  Dirac fermions .\cite{Ludwig}
The Hamiltonians for (2), (3) and (4)
involve only a single Pauli matrix at a time,  related to each other by unitary
transformations. Thus,  (2), (3) and (4) are
equivalent to each other and have $\nu \approx 7/3$.  Because unitary transformations
leave the identity matrix invariant,  the same argument implies
that (5), (6) and (7) are equivalent to each other and once again  $\nu \approx 7/3$.

The cases (8), (9) and (10) involve a pair of Pauli matrices and
are equivalent to each other. In (8) the Hamiltonian has a
discrete symmetry $\eta_{1}H \eta_{1}=-H$, often called a
particle-hole symmetry. The cases (9) and (10) have the same
discrete symmetry with respect to $\eta_{2}$ and $\eta_{3}$. The case
(10) has been analyzed by Hikami {\it et al.}\cite{Hikami} for a
spin degenerate nonrelativistic LLL.  When $g_{1}=g_{2}$, DOS diverges at the
band center and has two symmetrically located peaks away from it. The LD transition takes
place at these three distinct energies. Away from the band center
the LD transition has the exponent $\nu \sim 2.98$ and the transition
at the band center corresponds to a different exponent. If $g_{1} \neq g_{2}$,
the divergence of the DOS at the band center disappears, but the two
symmetrically placed peaks away from the band center still exist.
We find that the LD transition at these two energies hve continuously varying
exponents depending on the ratio $g_{2} / g_{1}$.

The cases (11), (12) and (13) are equivalent. The Hamiltonians in
these cases are respectively  the Hamiltonians for the cases (8), (9)
and (10),  augmented by the identity matrix corresponding to the
potential disorder. Potential disorder breaks the discrete
symmetry mentioned above, and there is no divergence
of the DOS at the band center. The DOS is still peaked at two
symmetrically placed energies away from the band center. The LD
transitions occur at energies away from the band center. If
$g_{0}$ is much smaller than the two remaining  coupling constants,
$\nu$ follows trends similar to (8), (9), and (10). If
$g_{0}$ is comparable or larger, we find $\nu \sim 7/3$.

In  (14) all three Pauli matrices are present. The
discrete symmetry of  (8), (9) and (10) are absent, and the
LD transitions take place at two symmetrically placed energies
away from the band center. When all the coupling constants are
equal, the exponent $\nu\sim 3.6$. Depending on the relative
strengths of the coupling constants  the exponents vary continuously. If any
particular coupling constant is significantly larger than
the rest, $\nu \sim 7/3$. By adding $g_{0}$ we obtain (15). If
$g_{0}$ is smaller than the rest, the
situation is similar to (14). If $g_{0}$ is larger than
the rest, $\nu \sim 7/3$.

\subsubsection{$m\ne 0$}

When $m\neq 0$, the LD transitions occur at two
symmetrically placed energies about the band center, and these energies are
greater than or equal to $m$. In the absence of internode
scattering, the transitions occur at $E=\pm m$ and the exponent
$\nu \sim 7/3$. If the strength of the intranode
scattering is larger than $m$, the bands at $\pm m$ overlap and
effectively correspond to the nodally degenerate case.

If $g_{0}=g_{3}=0$ and only one of the internode couplings is present,
the DOS diverges at $E=\pm m$  with an
exponent of 0.5 and is identically zero for $|E|<m$. The LD
transitions occur at $E=\pm m$ and have a continuously varying exponent.
When the disorder is strong compared to $m$, $\nu \sim 7/3$, and, in the
opposite limit, $\nu$ approaches unity.  If we include small intranode scattering the situation is similar.
If the intranode scattering strength is greater than the internode scattering, $\nu \sim 7/3$.

When $ g_{0}=g_{3}=0$ and both internode couplings are
present, the DOS  diverges at $E=\pm m$ with an exponent $\nu \sim
0.47$. However, the LD transitions occur at energies larger than
$|m|$. We have analyzed a case where $g_{1}=g_{2}$. The
exponent varies continuously. If the internode
scattering strength is larger than $m$,  $\nu \sim 3.8$,
and in the opposite limit $\nu$ approaches unity. This behavior is
stable against intranode scattering if its
strength is smaller than both $m$ and the internode
scattering. If intranode scattering strength is larger
than the internode scattering, $\nu \sim 7/3$.

\subsection{Roadmap}

Our paper is organized as follows: In Sec. II we describe the
Dirac fermion model. In Sec. III we describe
various possible disorders and their forms when projected to
the LLL.
 In sec IV we calculate the averaged density of states using supersymmetry.  In the
 sections V, VI, and VII we describe the numerical studies of
 the LD transition projected to the lowest Landau level. 
 Section  VIII is a brief concluding section. In the Appendix \ref{AppendixA} we
 provide some mathematical details of the density of states
 calculation. In Appendix \ref{AppendixB} we describe the recursive Green function
 technique used for numerical calculations and finally in Appendix \ref{AppendixC} we outline the procedure
 of data collapse involved in the finite size scaling of the localization length.

\section{Dirac fermions and Landau levels of graphene}
The low energy quasiparticles in graphene are well described by the
Lorentz invariant form as the sum over two inequivalent nodes (the Fermi velocity $v_{F}\approx 10^{6}\textrm{m/s}$)
\begin{equation}
H_{0}=-i\hbar v_{F} \int
d^{2}r\bar{\Psi}_{\sigma}\left(\gamma^{1}D_{x}+\gamma^{2}D_{y}\right)\Psi_{\sigma},
\end{equation}
where $\bar{\Psi}_{\sigma}=\Psi^{\dagger}_{\sigma}\gamma^{0}$ and the summation over spin $\sigma=\pm 1$ is understood.
The four component Dirac spinor $\Psi_{\sigma}= (\psi_{KA\sigma},i\psi_{KB\sigma},i\psi_{K'B\sigma},-i\psi_{K'A\sigma})$, where the component $\psi_{KA\sigma}$ is constructed by superposing Bloch functions close to one of the two inequivalent nodes $(K,K')$ of the Brillouin zone,  corresponding to one of the two sublattices $(A,B)$ of the hexagonal graphene lattice.   The notation  ${\bf D}=({\bm\partial}-i\frac{e}{c}{\bf A})$ stands for  the covariant derivative, $\bf A$ being the vector potential. The $\gamma$-matrices are defined by
$\gamma^{\mu} = (\tau_3, i \tau_1, i \tau_2) \otimes \eta_3$;  the Pauli matrix $\tau$ operates on the two components corresponding to the sublattice indices, and the Pauli matrix $\eta$ operates on the components corresponding to the nodal indices. To be explicit:
\begin{equation}
 \gamma^{0} =\begin{pmatrix}
\tau_3 & 0 \\ 0 &-\tau_3
\end{pmatrix},
\gamma^{1} =\begin{pmatrix} i \tau_1 & 0 \\ 0 &-i\tau_1
\end{pmatrix},
\gamma^{2} =\begin{pmatrix} i \tau_2 & 0 \\ 0 &-i\tau_2
\end{pmatrix}.
\end{equation}
To include a Zeeman term, we add
\begin{equation}
H_{z}=E_{z}\int d^{2}r \bar{\Psi}_{\sigma}\gamma^{0}\sigma_{3}^{\sigma\sigma'}\Psi_{\sigma'},
\end{equation}
where  $E_{z}=g\mu_{B}B$ is the Zeeman energy and
$\sigma_j$ is a Pauli matrix operating on the spin indices.
The Zeeman term breaks the $SU(2)$ symmetry of the spin space down
to $U(1)$. The energy eigenvalues of the Hamiltonian operator
\begin{equation}
{\hat{H}_{0}}=-i\hbar
v_{F} \gamma^{0}(\gamma^{1}D_{x}+\gamma^{2}D_{y})+E_{z}\sigma_{3},
\end{equation}
are well known:
\begin{eqnarray}
E_{ns\sigma}&=&s\sqrt{2n|eB| \hbar v_{F}^{2}/c}-\sigma E_{z},  n=0,1,2,...,
\end{eqnarray}
where $s=\pm1$ refer to the particle and the hole branches. In the presence of disorder Landau levels get broadened into a
band, and the amount of broadening depends on the strength of the
disorder. When the disorder is very strong, the half-width of the
broadened band can be larger than $E_{z}$, and experimentally this
corresponds to the spin degeneracy of the Landau bands. In the
spin degenerate situation, the observed filling factors is given
by  $\nu_{{\textrm f}}=4(n+\frac{1}{2})$.\cite{gusynin1,gusynin2}

The LLL wave function in the absence of the disorder in the symmetric gauge ${\bf A}=(-By/2, Bx/2,0)$
 can be written as 
\begin{equation}
{\cal U}(z,\bar{z})=e^{-z\bar{z}/4l_{B}^{2}}\begin{pmatrix} f_{1}(z)
\\ 0 \\ f_{2}(z) \\ 0 \end{pmatrix},
\end{equation}
where $eB> 0$.
The functions  $f_{1}(z)$ and $f_{2}(z)$ are holomorphic functions of the complex coordinates $z=x+iy$; $\bar z =x-iy$, and
$l_{B}=\sqrt{c/|eB|}$ is the magnetic length. Hence, in the zero mode, the first
and the second node have nonzero amplitudes coming only from the
sublattices A and B respectively. 

Two distinct onsite energies on the two sublattices correspond to a charge
 density modulation at the lattice scale. As a result,  the particle and the hole branches  acquire an energy gap. When linearized
 about the inequivalent nodes, this energy gap appears as a parity preserving mass
 of the Dirac fermions. To be explicit, the linearized hamiltonian will have two
 new terms: the chemical potential term $[(V_{A}+V_{B})/2]\bar{\Psi}\gamma^{0}\Psi$
 and the mass term $[(V_{A}-V_{B})/2]\bar{\Psi}\Psi$, where $V_{A}$ and $V_{B}$ are the
 site energies at the sublattices $A$ and $B$. 

Although the non-interacting quasiparticles
are massless in the absence of site modulation, they can acquire a parity conserving mass
due to interaction effects. This spontaneous symmetry breaking
is facilitated by the presence of the magnetic field, a phenomenon
known as ``magnetic catalysis'' of chiral symmetry
breaking.\cite{gusynin4, Khveshchenko} The effect has
been argued to be the reason behind the  quantum hall plateaus at $\nu_{f}=0,\pm1$ observed in
strong magnetic fields.\cite{gusynin3, herbut} Though it is beyond
the scope of the present paper to consider
electronic interactions, we will pay some attention to the
noninteracting problem with a finite mass. Our philosophy is
to analyze the consequences of having a mass (possible in
an interacting theory) on the LD transition. So, we shall include the term
$m\Psi^{\dagger}\gamma^{0}\Psi$ in the
effective Hamiltonian to examine the effect of mass.
In the presence of such a mass term, the nodal degeneracy of $E_{0,s,\sigma}=-\sigma E_{z}$ is removed and it splits into four levels $E_{0,1,\sigma}=m-\sigma E_{z}$ and $E_{0,-1,\sigma}=-m-\sigma E_{z}$. Each of these levels has the degeneracy $ \frac{|eB|}{2\pi c}$. If the applied chemical potential is smaller than $|E_{z}-m|$, there will be a plateau at $\nu_{\textrm f}=0$. If $|E_{z}-m|<|\mu|<E_{z}+m$, $\nu_{\textrm f}=\pm1$ plateaus will appear depending on the sign of $\mu$.\cite{gusynin4, alicea} Next possible values of quantized plateaus are $\nu_{\textrm f}=\pm2$. The introduction of the mass term does not, however, lift the nodal degeneracy of the higher Landau levels, and the energy levels $E_{n\geq1,s,\sigma}=s \sqrt{m^{2}+2n|eB| \hbar v_{F}^{2}/c}-\sigma E_{z}$ has the degeneracy $\frac{|eB|}{\pi c}$. Therefore, when a mass is included, quantized plateaus appear at $\nu_{\textrm f}=0,\pm1,\pm 2q$, where $q$ is an integer.

\section{Randomness}
There are many sources of disorder in graphene : vacancies, interstitials, substrate disorder and lattice distortions due to dislocations. In principle there could  also be random spin-orbit coupling. However, due to the small atomic mass of carbon, spin-orbit coupling is very weak compared to other energy scales.  For simplicity, we shall primarily be interested in the spin polarized limit and ignore the random spin-orbit coupling.

Point defects and substrate disorder can be described by introducing random site energies in the tight binding model. In the presence of substrate disorder there can also be a random modulation of the charge densities between the two sublattices. These effects can be described by a random chemical potential $V_{0}({\bf r})\bar{\Psi}\gamma^{0}\Psi$ and a random mass  $V_{3}({\bf r})\bar{\Psi}\Psi$ in the continuum limit. 

Because  true long range crystalline order is not possible in two dimensions at any finite temperature, topological defects, dislocations and disclinations will be present.  Effects of these topological defects will result in random hopping amplitudes $\delta t_{AB}$ and hence intranode as well as internode scattering. However, these scattering processes will take place between states on different sublattices.   The follwing two bilinears, $V_{2}({\bf r})\bar{\Psi} \gamma^{3}\Psi$ and $V_{1}({\bf r})\bar{\Psi}\gamma^{5} \Psi$,  describe the internode scattering terms arising from random hopping. The two mutually anticommuting matrices,
 \begin{equation}
 \gamma^{3} =i\begin{pmatrix}
0 & I \\ I &0
\end{pmatrix},  \gamma^{5} =\begin{pmatrix}
0 & -I\\ I &0
\end{pmatrix},
\end{equation}
also anticommute with $\gamma^{\mu}$; $I$ is the identity matrix.

 In the continuum limit, the most general impurity Hamiltonian is a $4\times 4$ matrix:
\begin{equation}
H^{\textrm{imp}}=\int d^2r \Psi^{\dagger}_{\sigma}\begin{pmatrix} D_{11}(\bf{r}) &D_{12}(\bf{r}) \cr D_{21}(\bf{r})
&D_{22}(\bf{r})
\end{pmatrix}\Psi_{\sigma},
\end{equation}
where $D_{ij}(\bf{r})$ are $2\times2$ matrices. Here
$D_{11}=D_{11}^{\dagger}$ and $D_{22}=D_{22}^{\dagger}$ represent
intranode scattering at nodes 1 and 2 respectively;  $D_{12} =
D_{21}^{\dagger}$ represent internode scattering. 

After projecting to the LLL, the disorder matrix reduces to a
$2\times2$ matrix and can be represented by the Pauli matrices $\eta$.  The most
general disorder matrix projected to LLL then takes the form
\begin{equation}
{\hat{H}}^{\textrm{imp}}_{\textrm{LLL}}=\sum_{j=0}^{3}V_{j}({\bf r})\eta_{j}. \label{eq:impLLL},
\end{equation}
where we have denoted the $I$ matrix by $\eta_{0}$. 

\section{Average Density of States}
Using a four component bosonic spinor $\phi$ and a four component Grassmann spinor $\chi$
the average retarded
 Green function for a noninteracting problem can be written as
\begin{eqnarray}
\overline{G}^{R}(E;r,r^{'})=- i \prod_{j=0}^{3}\int \mathcal{D}[\phi^{*}]\mathcal{D}[\phi]\mathcal{D}[\chi^{*}]\mathcal{D}[\chi]\nonumber \\
\mathcal{D}[V_{j}]P[V_{j}]\phi^{*}(r)\phi(r^{'}) e^{S^{R}} ,
\end{eqnarray}
where $P[V_{j}]$ is the probability distribution of $V_{j}$ and 
\begin{eqnarray}
S^{R}=i \int d^{2}r \bigg[\phi^{\dagger}(E-\hat{H}_{0}-\hat{H}^{\textrm{imp}} + i\delta)\phi +\nonumber \\ 
\chi^{\dagger}(E-\hat{H}_{0}-\hat{H}^{\textrm{imp}} + i\delta)\chi \bigg].
\end{eqnarray}
The average density of states is given by
\begin{equation}
\overline{\rho}(E)=- \frac{1}{\pi}{\textrm{Im}} \; \overline{G}^{R}(E;r,r).
\end{equation}
After performing the disorder averages we can write
\begin{eqnarray}
\overline{G}^{R}(E;r,r^{'})=- i\int \mathcal{D}[\phi^{*}]\mathcal{D}[\phi]\mathcal{D}[\chi^{*}]\mathcal{D}[\chi]\nonumber \\
\phi^{*}(r)\phi(r^{'}) e^{\mathcal{A}^{R}} ,
\end{eqnarray}
where the action $\mathcal{A}^{R}$ involves interactions among the fields generated by the disorder averaging procedure.
After projection to the LLL, the action $\mathcal{A}^{R}$ can be expressed in terms of a two-component holomorphic bosonic spinor
\begin{equation}
\phi(z)=\begin{pmatrix} v_{1} (z)\\ v_{2}(z)\end{pmatrix},
\end{equation}
and a two component holomorphic Grassmann spinor
\begin{equation}
\chi(z)=\begin{pmatrix} w_{1} (z)\\ w_{2}(z)\end{pmatrix}.
\end{equation}
 In terms of these fields the action is given by
\begin{eqnarray}
\mathcal{A}^{R}&=&\mathcal{A}^{R}_{f}+\sum_{j=0}^{3}\mathcal{A}^{D}_{j}, \nonumber \\
\mathcal{A}^{R}_{f}&=&i \epsilon \int d^{2}z e^{-z\bar{z}/2l_{B}^{2}}
(\phi^{\dagger}\phi+\chi^{\dagger}\chi),\nonumber \\
\mathcal{A}^{D}_{j}&=&\int d^{2}z h_{j}\bigg[e^{-z\bar{z}/2l_{B}^{2}}(\phi^{\dagger}\eta_{j}\phi+\chi^{\dagger}\eta_{j}\chi)\bigg],
\end{eqnarray}
where $\epsilon=E+i\delta$ and 
\begin{equation}
h_{j}(\kappa)=\ln \bigg(\int e^{-i\kappa V_{j}}P[V_{j}] \mathcal{D}V_{j} \bigg),
\end{equation}
is the effective interaction of the fields generated by the averaging over the random variable $V_{j}$. For the Cauchy distribution, defined by
 \begin{equation}
P[V_{j}(\vec{r})]=\frac{g_{j}}{\pi}\frac{1}{g_{j}^{2}+V_{j}^{2}(\vec{r})},
\end{equation}
we have
\begin{equation}
h_{j}(\kappa)=-g_{j}|\kappa|.
\end{equation}
If the disorder distribution is Gaussian white noise, defined by
\begin{equation}
P[V_{j}(\vec{r})]={\cal N} \exp[-\frac{1}{2g_{j}}\int d^{2}r V_{j}^{2}(\vec{r})],
\end{equation}
we get
\begin{equation}
h_{j}(\kappa)=-\frac{1}{2}g_{j}\kappa^{2}.
\end{equation}

The above action is  invariant under the translation followed by a gauge transformation. Due to this invariance, the spatial dependence of the average retarded Green function is same as the spatial dependence of the pure system's Green function
\begin{equation}
G_{pure}^{R}(E, z_{1}, z_{2})=\frac{\exp[-(|z_{1}|^{2}+|z_{2}|^{2}-2z_{1} \bar{z}_{2}]}{2\pi l_{B}^{2}(E+i\delta)}.
\end{equation}
The disorder averaged Green function can be written as
\begin{equation}
\overline{G}^{R}(E, z_{1},
z_{2})=C(E+i\delta,g_{j})\exp[-(|z_{1}|^{2}+|z_{2}|^{2}-2z_{1}
\bar{z}_{2}],
\end{equation}
where $g_{j}$'s are coupling constants of various types of
disorder and $C(E+i\delta,g_{j})$ is a gauge invariant
proportionality constant which depends on the energy and disorder
strengths. This gauge invariant proportionality constant is what
we need to calculate to find the average density of states.

For the calculation of the average Green function's dependence on the energy and disorder coupling constants we introduce two new Grassmann variables $\theta$ and $\bar{\theta}$ and enlarge the Euclidean coordinate space into a superspace of coordinates $(x,y,\theta,\bar{\theta})$. Integrals over the Grassmann coordinates are normalized as $\pi \int d\theta d\bar{\theta}\bar{\theta}\theta=1$. The norm of a coordinate vector is defined as $x^{2}+y^{2}+\bar{\theta}\theta$. This norm is invariant under the superspace rotations. In addition to the
ordinary rotations in the Euclidean subspace and the symplectic transformations in the Grassmann subspace,
the superspace rotations involve transformations which mix $(x,y)$ and $(\theta,\bar{\theta})$
in the following manner:
\begin{eqnarray}
\vec{r}&\rightarrow& \vec{r}+2\vec{l}_{1}\Omega \theta +2\vec{l}_{2}\Omega \bar{ \theta} \nonumber \\
\theta&\rightarrow& \theta +4(\vec{l}_{2}\cdot \vec{r})\Omega \nonumber \\
\bar{\theta}&\rightarrow& \bar{\theta}-4(\vec{l}_{1}\cdot \vec{r})\Omega.
\end{eqnarray}
In the above set of transformations $\vec{l}_{1,2}$ are two
arbitrary Euclidean vectors and $\Omega$ is a Grassmann number. We
also define two holomorphic superfields and their conjugates
as
\begin{eqnarray}
\Phi(z,\theta)&=&\phi(z)+\frac{\theta}{\sqrt{2} l_{B}}\chi(z), \nonumber \\
\bar{\Phi}(z,\theta)&=&\phi^{\dagger}(z)+\frac{\chi^{\dagger}(z)}{\sqrt{2} l_{B}}\bar{\theta}.
\end{eqnarray}

In terms of these superfields the pure part of the action can
be expressed as
\begin{equation}
\mathcal{A}_{f}^{R}=2i\epsilon\pi l_{B}^{2}\int d^{2}z d\theta
d\bar{\theta} e^{-(z\bar{z}+\theta
\bar{\theta})/2l_{B}^{2}}\bar{\Phi}\Phi,
\end{equation}
which is manifestly invariant under superspace rotations. After
the disorder contributions to the action are expressed in terms of
these new superfields, we have to demonstrate these to be
invariant under superspace rotations. In order to be
supersymmetric $\mathcal{A}^{D}_{j}$'s have to be local in the
supercoordinate space and this is only possible if they do not
involve any quartic fermionic interactions. We note that
\begin{eqnarray}
&&h_{j}\bigg[e^{-z\bar{z}/2l_{B}^{2}}(\phi^{\dagger}\eta_{j}\phi+\chi^{\dagger}\eta_{j}\chi)\bigg]\nonumber \\
&&=h_{j}\bigg[e^{-z\bar{z}/2l_{B}^{2}}\phi^{\dagger}\eta_{j}\phi\bigg]\nonumber \\
&&+h^{'}_{j}\bigg[e^{-z\bar{z}/2l_{B}^{2}}\phi^{\dagger}\eta_{j}\phi\bigg]e^{-z\bar{z}/2l_{B}^{2}}\chi^{\dagger}\eta_{j}\chi \nonumber
\\
&&+\frac{1}{2}h^{''}_{j}\bigg[e^{-z\bar{z}/2l_{B}^{2}}\phi^{\dagger}\eta_{j}\phi\bigg]e^{-z\bar{z}/l_{B}^{2}}(\chi^{\dagger}\eta_{j}\chi)^{2},
\end{eqnarray}
where $h^{'}_{j}$ and $h^{''}_{j}$ correspond to the first and second derivatives of $h_{j}$ with respect to its argument. The Taylor series truncates at the quadratic order as the higher powers of $\chi^{\dagger}\eta_{j}\chi$ are identically zero according to the anticommutation rules. We also note that 
$(\chi^{\dagger}\eta_{j}\chi)^2=-2w_{1}^{\dagger}w_{1}w_{2}^{\dagger}w_{2}$ for $j=1,2,3$ and $(\chi^{\dagger}\eta_{0}\chi)^2=2w_{1}^{\dagger}w_{1}w_{2}^{\dagger}w_{2}$. If $h^{''}_{j}$ does not vanish we get four-fermion interactions. 

If there were one bosonic and one Grassmann fields instead of spinors, as in the problem solved by  of Br\'ezin {\it et al.},~\cite{Brezin} no four fermionic terms would be generated, and the action for an arbitrary disorder distribution would be local in the superspace coordinates.  For the case under consideration, such a simplification is not possible in general.  However, for Cauchy distribution the disorder averaged action is quadratic and can be made manifestly supersymmetric. Thus,  the calculation of the DOS reduces to a calculation of a zero dimensional field theory  over two complex bosonic fields.

\begin{widetext}
\subsection{Cauchy Distribution}

\subsubsection{$m=0$}

The action is given by
\begin{equation}
\mathcal{A}^{R}=\int d^{2}z e^{-z\bar{z}/2l_{B}^{2}}\bigg[i \epsilon(\phi^{\dagger}\phi+\chi^{\dagger}\chi )             -\sum_{j=0}^{3}g_{j}|\phi^{\dagger}\eta_{j}\phi+\chi^{\dagger}\eta_{j}\chi|\bigg].
\end{equation}
Using the superfields $\Phi$ and $\bar{\Phi}$ the action can
be written as,
\begin{eqnarray}
\mathcal{A}^{R}=&2\pi l_{B}^{2}&\int d^{2}z d\theta d\bar{\theta}
e^{-(z\bar{z}+\theta \bar{\theta})/2l_{B}^{2}}\bigg[i \epsilon
\bar{\Phi}\Phi
              -\sum_{j=0}^{3}g_{j}|\bar{\Phi}\eta_{j}\Phi|\bigg],
\end{eqnarray}
which is manifestly invariant under rotation and magnetic translation in superspace. Because of this symmetry,
the DOS can be reduced to a simple expression involving integrals over
two ordinary complex variables. Expressed in terms of two radial and two angular variables, it is
\begin{eqnarray}
\overline{\rho}(E)=\frac{1}{2\pi^{2} l_{B}^{2}}&&{\textrm{Im}}
\frac{\partial}{\partial \epsilon}\ln \bigg\{
\int_{0}^{\infty}d(r_{1}^{2}/2)\int_{0}^{\infty} d(r_{2}^{2}/2)
\int_{0}^{2\pi} d\alpha_{1} \int_{0}^{2\pi} d\alpha_{2}
                      \exp\bigg[i \epsilon(r_{1}^{2}+r_{2}^{2})-g_{0}|r_{1}^{2}+r_{2}^{2}|\nonumber \\
                      &&-g_{3}|r_{1}^{2}-r_{2}^{2}|-2g_{1}r_{1}r_{2}|\cos(\alpha_{1}-\alpha_{2})|-2g_{2}r_{1}r_{2}|\sin(\alpha_{1}-\alpha_{2})|\bigg]\bigg\}.
\label{eq:dencauch}
\end{eqnarray}

For simplicity, consider the cases where we keep only one of the internode scattering, or the random mass term, along with the potential disorder. We get:

$(i)\; g_{1}=g_{2}=0$
\begin{equation}
\overline{\rho}(E)=\frac{1}{2\pi^{2}l_{B}^{2}}\bigg[\frac{g_{0}}{g_{0}^{2}+E^{2}}+\frac{g_{0}+g_{3}}{(g_{0}+g_{3})^{2}+E^{2}}\bigg],
\end{equation}
$(ii) \; g_{3}=g_{2}=0$,
\begin{equation}
\overline{\rho}(E)=\frac{1}{2\pi^{2}l_{B}^{2}}\bigg[\frac{g_{0}}{g_{0}^{2}+E^{2}}+\frac{g_{0}+g_{1}}{(g_{0}+g_{1})^{2}+E^{2}}\bigg].
\end{equation}
The answer for the case
$(iii)\; g_{3}=g_{1}=0$ is identical to the case $(ii)$. The DOS
obtained for these three cases are identical, as the Hamiltonian involves only  one Pauli matrix at a time,
and these matrices are related by unitary transformations.

Consider now
$g_{1}=g_{2}=g_{IN}$ and $g_{3}=0$. We obtain, defining by $I$ the expression within the curly parenthesis in Eq.~\ref{eq:dencauch},
\begin{equation}
I=-\frac{\pi}{(a^{2}-2g_{IN}^{2})}\bigg[\pi-4\frac{g_{IN}}{\sqrt{a^{2}-g_{IN}^{2}}}\tan^{-1}\bigg(\frac{g_{IN}}{\sqrt{a^{2}-g_{IN}^{2}}}\bigg)-2\sqrt{2}\pi
\frac{g_{IN}}{a}+2\pi
\frac{g_{IN}}{\sqrt{a^{2}-g_{IN}^{2}}}\bigg],
\label{eq:integral1}
\end{equation}
where $a=g_{0}-i\epsilon$. The details of the evaluation of the
multiple integrals are provided in the Appendix~\ref{AppendixA}. The expression
for the DOS obtained from this expression is lengthy and not very
illuminating, but it is important to note that because of the presence of
the term $\tan^{-1}\left(g_{IN}/\sqrt{a^{2}-g_{IN}^{2}}\right)$,  we obtain a
$\ln E$ divergence at the band
center when $g_{0}=0$. Based on symmetry, similar behavior will be obtained when a combination of two Pauli
matrices are considered. This should be contrasted with the $(\ln E)^{2}$ divergence obtained by Hikami {\em et al.}~\cite{Hikami}

\subsubsection{$m\ne 0$}
When the fermion is massive, we will ignore the mass disorder
part. The density of states is given by
\begin{eqnarray}
\overline{\rho}(E)=\frac{1}{2\pi^{2} l_{B}^{2}}Im
\bigg
(\frac{\partial}{\partial
\epsilon_{1}}+\frac{\partial}{\partial \epsilon_{2}}\bigg)&&\ln\bigg\{
\int_{0}^{\infty}d(r_{1}^{2}/2)\int_{0}^{\infty}  d(r_{2}^{2}/2)
\int_{0}^{2\pi} d\alpha_{1}\int_{0}^{2\pi} d\alpha_{2}
                      \exp\bigg[i \epsilon_{1}r_{1}^{2}+i\epsilon_{2}r_{2}^{2}\nonumber \\
                      &&-g_{0}|r_{1}^{2}+r_{2}^{2}|-2g_{1}r_{1}r_{2}|\cos(\alpha_{1}-\alpha_{2})|-2g_{2}r_{1}r_{2}|\sin(\alpha_{1}-\alpha_{2})|\bigg]\bigg\},
\label{eq:massdencauch}
\end{eqnarray}
where $\epsilon_{1,2}=E\pm m+i\delta$. Again if we take only one
of the internode scattering terms ($g_{2}=0$) for simplicity, the
expression within the curly parenthesis  in Eq.~\ref{eq:massdencauch}, $I$, becomes
\begin{equation}
I=-\frac{\pi^2}{ab+g_{1}\sqrt{ab}}
 \label{eq:int1}
\end{equation}
The density of states is then given by
\begin{eqnarray}
\overline{\rho}(E)=\frac{1}{4\pi^{2}
l_{B}^{2}}\sum_{\sigma=\pm1}\frac{g_{0}}{g_{0}^{2}+(E+\sigma m)^{2}}+\frac{1}{2\pi^{2}
l_{B}^{2}}\frac{g_{0}(R\cos\beta
+g_{1}\sqrt{R}\cos\frac{\beta}{2})+E(R\sin\beta+g_{1}\sqrt{R}\sin\frac{\beta}{2})}{R^{2}+g_{1}^{2}R+2g_{1}R^{\frac{3}{2}}\cos\frac{\beta}{2}},
\end{eqnarray}
where $R=\sqrt{(g_{0}^{2}+m^{2}-E^{2})^{2}+4g_{0}^{2}E^{2}}$ and $\tan\beta=(2g_{0}E/(g_{0}^{2}+m^{2}-E^{2}))$.  The above expression takes particularly simple form
when $g_{0}=0$. It becomes
\begin{eqnarray}
\overline{\rho}(E^{2}>m^{2})&=&\frac{1}{4\pi^{2} l_{B}^{2}}\bigg[\delta(E+m)+\delta(E-m)+\frac{2Eg_{1}}{(E^{2}-m^{2}+g_{1}^{2})\sqrt{E^{2}-m^{2}}}\bigg],\nonumber \\
\bar{\rho}(E^{2}<m^{2})&=&0.
\end{eqnarray}

If both internode scatterings are present and
$g_{1}=g_{2}=g_{IN}$, the integral is given by
\begin{equation}
I=-\frac{\pi}{(ab-2g_{IN}^{2})}\bigg[\pi-4\frac{g_{IN}}{\sqrt{ab-g_{IN}^{2}}}\tan^{-1}\bigg(\frac{g_{IN}}{\sqrt{ab-g_{IN}^{2}}}\bigg)-2\sqrt{2}\pi
\frac{g_{IN}}{\sqrt{ab}}+2\pi
\frac{g_{IN}}{\sqrt{ab-g_{IN}^{2}}}\bigg]
\label{eq:int2}
\end{equation}
The expression for the DOS is tedious. However for $g_{0}=0$,the
feature that the DOS is zero for $E^{2}<m^{2}$ is still valid. In
this case for energies close to $\pm m$,  $\overline{\rho}(E)\sim
\ln|E-m|/\sqrt{|E-m|}$.

\section{Hall plateau in the lowest Landau level}

Similar to the method described in
Ref.~\onlinecite{Huckestein:1995}, we generate the matrix elements
of the Dirac Hamiltonian after projecting to the lowest Landau
level. In our problem, the element $\langle k|H|k'\rangle$  itself
is a $2\times 2$ matrix:
\begin{equation}
        \langle k|H|k'\rangle=\int dx dy \psi_k^{*}(x,y)[H^{\textrm{imp}}_{\textrm{LLL}}(x,y)+m\eta_3]\psi_{k'}(x,y)
    =m\eta_3\delta_{k,k'}+V(k,k'),
\end{equation}
\end{widetext}
where $\psi_k(x,y)$ is the lowest Landau level
wave function in the Landau gauge. We choose all the $V_{j}$'s to follow independent Gaussian white noise distributions such that 
$\overline{V_{j}(x,y)V_{j'}(x',y')}=g_j^2\delta_{j,j^{'}}\delta(x-x')\delta(y-y')$. Then the elements of the $2\times2$ matrix $V(k,k')$ can be computed
explicitly---for example,
\begin{widetext}
\begin{equation}
    V(k,k')_{11}=\frac{1}{\sqrt{\pi
    L_y}}\mathrm{e}^{-\frac{1}{4}l_B^2(k-k')^2}\int\mathrm{d}\xi~\mathrm{e}^{-\xi^2}[g_0u_0(l_B\xi+\frac{k+k'}{2}l_B^2,k'-k)
    +g_3u_3(l_B\xi+\frac{k+k'}{2}l_B^2,k'-k)],
    \label{exa_mat_ele}
\end{equation}
\end{widetext}
where $u_j(x,k)$ is a complex random variable defined to be the
Fourier transforms of $V_{j}(x,y)$ along the $y$ direction normalized by the width $g_j$, namely:
\begin{equation}
    u_j(x,k)=\frac{1}{g_j\sqrt{L_y}}\int dy\;V_{j}(x,y)e^{iky}.
\end{equation}
Because each of the disorder fields has 
zero correlation length, and there are no correlations between them,
\begin{equation}
    \overline{u_i(x,k)u_j(x',k')}=\delta_{i,j}\delta(x-x')\delta(k+k')
    \label{cor_u}
\end{equation}

It is
straightforward to compute the statistical properties of the matrix
elements. The averages are:
\begin{equation}
    \overline{V(k,k')_{i,j}}=0\mbox{\quad:\quad} i,j=1,2
\end{equation}
As to correlations, the only non-vanishing pairs are:
\begin{widetext}
\begin{equation}\label{correlation}
\begin{split}
    \overline{V(k_1,k_2)_{11}V(k_3,k_4)_{11}}&=\overline{V(k_1,k_2)_{22}V(k_3,k_4)_{22}} =       \frac{g_0^2+g_3^2}{\sqrt{2\pi}L_y}\exp{[-\frac{l_B^2}{2}((k_1-k_2)^2+(k_4-k_1)^2)]}\delta_{k_1-k_2,k_4-k_3}\\
    \overline{V(k_1,k_2)_{11}V(k_3,k_4)_{22}}&=\overline{V(k_1,k_2)_{22}V(k_3,k_4)_{11}}  =      \frac{g_0^2-g_3^2}{\sqrt{2\pi}L_y}\exp{[-\frac{l_B^2}{2}((k_1-k_2)^2+(k_4-k_1)^2)]}\delta_{k_1-k_2,k_4-k_3}\\
    \overline{V(k_1,k_2)_{12}V(k_3,k_4)_{12}}&=\overline{V(k_1,k_2)_{21}V(k_3,k_4)_{21}}   =     \frac{g_1^2-g_2^2}{\sqrt{2\pi}L_y}\exp{[-\frac{l_B^2}{2}((k_1-k_2)^2+(k_4-k_1)^2)]}\delta_{k_1-k_2,k_4-k_3}\\
    \overline{V(k_1,k_2)_{12}V(k_3,k_4)_{21}}&=\overline{V(k_1,k_2)_{21}V(k_3,k_4)_{12}}   =     \frac{g_1^2+g_2^2}{\sqrt{2\pi}L_y}\exp{[-\frac{l_B^2}{2}((k_1-k_2)^2+(k_4-k_1)^2)]}\delta_{k_1-k_2,k_4-k_3}\\
\end{split}
\end{equation}
\end{widetext}

For numerical implementation, we discretize and use the integer $l$ to  label the $x$ coordinate. We then
generate a set of complex random variables $u_j(l,k)$,
that are $\delta$-correlated as in (\ref{cor_u}).
Finally, we approximate the integrals by sums. Explicitly, the
matrix  elements are
\begin{widetext}
\begin{equation}
\begin{split}
    V(k,k+k')_{11}&=\frac{e^{-a^2 k'^2}}{\sqrt{M
    A}}\sum_j \left[g_0 u_{0}(2k+j,k')+g_3 u_{3}(2k+j,k')\right]e^{-a^2
    j^2}\\
    V(k,k+k')_{22}&=\frac{e^{-a^2 k'^2}}{\sqrt{M
    A}}\sum_j \left[ g_0 u_{0}(2k+j,k')-g_3 u_{3}(2k+j,k')\right]e^{-a^2
    j^2}\\
    V(k,k+k')_{12}&=\frac{e^{-a^2 k'^2}}{\sqrt{M
    A}}\sum_j \left[ g_1 u_{1}(2k+j,k')-ig_2 u_{2}(2k+j,k')\right]e^{-a^2
    j^2}\\
    V(k,k+k')_{21}&=\frac{e^{-a^2 k'^2}}{\sqrt{M
    A}}\sum_j \left[ g_1 u_{1}(2k+j,k')+ig_2 u_{2}(2k+j,k')\right]e^{-a^2
    j^2}
\end{split}
\end{equation}
\end{widetext}
with $A=\sum_j \mathrm{e}^{-2a^2 j^2}$ and $a^2=\pi/2M^2$. Here $M$
is the length of the system in the $y$ direction, the unit being
$\sqrt{2\pi}l_B$, that is, $M=L_y/\sqrt{2\pi}l_B$,  chosen to be an
integer. The integers  $k$ and $k'$ label the wave vectors. Since the
matrix elements decay exponentially, we can neglect them for $k'>2M$.
A  cutoff is also necessary for  the recursive Green's function
technique\cite{Huckestein:1995} that we use.

We compute the density of states $\rho(E)$ by directly diagonalizing
the Hamiltonian. We have
checked that $\rho(E)$ is independent of  $M$, for sufficiently large $M$; $M=32$ seems to be sufficient;
the total number of momentum states $N_k$ is chosen to be $1000$,
which is half the dimension of the Hamiltonian matrix to be
diagonalized, as there are two fermions for each $k$. Typically,  an
average over $100$ disorder realizations is used.

The recursive Green's function technique, similar to that in
Ref.~\onlinecite{Huckestein:1995}, is used to explore the
localization properties. The details are
described in Appendix~\ref{AppendixB}. We first compute the localization lengths
for a finite system, $\lambda_M(E_i)$, at a set of energies,
$\{E_i\}_{i=1}^{N_E}$, in systems with transverse dimensions
$\{M_j\}_{j=1}^{N_M}$. Since there are two types of fermions, in general there can be two distinct localization lengths;
however,  in most  cases discussed below, they
are identical within our numerical accuracy, and we will not generally distinguish them.
Assuming finite-size scaling, $\lambda_M(E)/M=f(M^{1/\nu}(E-E_c))$,
where $f(x)$ is a universal function, the data is collapsed to
obtain the localization length exponent $\nu$, and the critical
energy $E_c$. Strictly, scaling
holds only for large enough systems in the vicinity of
critical energy. Here the energies $\{E_i\}$ are chosen
close to the critical energy, $E_c$, and the validity of the scaling
law is verified by the success of data collapse. For the details of
the  procedure involving data collapse, see Appendix~\ref{AppendixC}.

The numerical calculations about the localization properties were
mostly performed for a quasi-one dimensional system with the transverse
dimensions $M= 8,16,32,64$. The total number of momentum states is
$N_k=5\times 10^4$. Because of the $2\times 2$ character of the
Hamiltonian matrix elements, the numerical calculations are more demanding than those
in Ref.~\onlinecite{Huckestein:1995}. The data are typically averaged over
100 disorder configurations to reduce fluctuations. Energies
$\{E_i\}$ were chosen close to the critical energy and measured in
units of $2(\sum_{j=0}^{3}g_j^2)^{1/2}$ similar to
Ref.~\onlinecite{Huckestein:1995}.

Our program is also validated by the case  $g_0=0.5$,
$g_1=g_2=g_3=0$, and $m=0$. In this case, the two types of fermions
are independent. Because the LLL wave function is identical to the
non-relativistic one, the properties should be the same as in
Ref.~\onlinecite{Huckestein:1995}. Numerical computations show a
single peak in the density of states and a localization length
exponent of $\nu=2.41\pm 0.08$; both agree well with the previous
results.

\section{LD transition for the massless case}

\subsection{One disorder field}

Consider first the cases where only one type of disorder has nonzero
strength. Numerically, we considered (1) $g_3=0.5$, $g_0=g_1=g_2=0$, (2)
$g_1=0.5$, $g_0=g_2=g_3=0$, and (3)$g_2=0.5$, $g_0=g_1=g_3=0$. In
all of these cases, the delta function density of states in the pure
system are broadened into a simple bell shape function due to
disorder.  The  results of successful
data collapse, not shown here,
yield critical exponents $\nu=2.46\pm 0.09$, $\nu=2.48\pm 0.11$ and
$\nu=2.45\pm 0.08$ for cases (1), (2), and (3), respectively, that is, they are
the same within the error bars.

The  critical exponents  are all equal to that  of  
single type of fermions subject to potential disorder.
This can be understood as follows. The original Hamiltonian matrix is in the basis
$\{|k_1,1\rangle, |k_1,2\rangle, |k_2,1\rangle, |k_2,2\rangle,
\ldots\}$, where $1$ and $2$ label the type of fermions. If we
reorder the basis as $\{|k_1,1\rangle, |k_2,1\rangle, \ldots,
|k_1,2\rangle, |k_2,2\rangle, \ldots\}$, the Hamiltonian becomes
a $2\times 2$ block matrix with diagonal blocks representing
intranode parts, and the off-diagonal blocks representing internode
scatterings. Explicitly, it is in the form:
\begin{equation}\label{hamiltonian}
    H=\left(
        \begin{array}{cc}
          g_0U_0+g_3U_3+mI & g_1U_1-ig_2U_2 \\
          g_1U_1+ig_2U_2 & g_0U_0-g_3U_3-mI \\
        \end{array}
      \right)
\end{equation}
where $U_i$, $i=0,\ldots,3$ are statistically independent random Landau matrices, and $I$ is the
identity matrix. The case with only $g_3$ nonzero, has the
same structure, and hence same statistical properties, as the
case when only $g_0$ is nonzero. When only $g_1$ is nonzero,
we can, by a unitary transformation given by,
\begin{equation}\label{rot_y}
    T=\mathcal{B}\left(
        \begin{array}{cc}
          I & I \\
          I & -I \\
        \end{array}
      \right),
\end{equation}
where  $\mathcal{B}$ is a normalization constant, 
bring the Hamiltonian back to the block diagonal  form, resulting in
a structure corresponding to two types of independent
fermions in the presence of mass disorder. Thus, the critical exponent is the same as the case when only $g_0$ is non-zero. The
same argument also applies to  the case when only  $g_2$ is non-zero.

\subsection{Two disorder fields}

If one of the two disorder  fields is
$V_0(x,y)$, and another is  $V_{1}(x,y)$, or $V_2(x,y)$, or
$V_3(x,y)$,  an appropriate
unitary transformation about an axis by $\pi/2$ will map one possible case to another. For instance,
the transformation  in Eq.~(\ref{rot_y}) will transform the case with $V_0(x,y)$ and
$V_2(x,y)$ to $V_{0}(x,y)$ and $V_3(x,y)$.  It is therefore
sufficient to consider only the case with just $V_0(x,y)$ and $V_3(x,y)$.
However, from Eq.~ (\ref{hamiltonian}), the Hamiltonian is
block diagonal, and the blocks $g_0U_0+g_3U_3$ and
$g_0U_0-g_3U_3$ are statistically equivalent to a new block
$\sqrt{g_0^2+g_3^2}U$, with $U$ a new random matrix
satisfying the same statistical properties as $U_i$'s; see
Eq.~(\ref{correlation}). That is, the Hamiltonian for $g_0\ne 0$ and
$g_3\ne 0$ is statistically the same as that corresponding to  a  potential
disorder 
$\tilde{g_0}=\sqrt{g_0^2+g_3^2}$.

Our numerical computations confirm this argument. The data collapse was found to be successful,
assuming $E_{c}=0$, and the 
critical exponents are $\nu=2.45\pm 0.06$ for 
$g_0=g_3=0.5$, $g_1=g_2=0$, and $\nu=2.45\pm 0.11$ for
$g_0=g_1=0.5$, $g_2=g_3=0$.

Next, we choose two disorder fields from $V_1(x,y)$, $V_2(x,y)$, and
$V_3(x,y)$. There are three possible combinations. In fact,
these three cases are not independent; we can map
one  case to another by an appropriate unitary
transformation corresponding to a rotation by $\pi/2$ about a
certain axis. Therefore, it is sufficient to consider only one of the
three cases; for example, let us choose $V_3(x,y)$ (mass disorder), and
 $V_1(x,y)$ (internode coupling).

\begin{figure}
\includegraphics[scale=1.2]{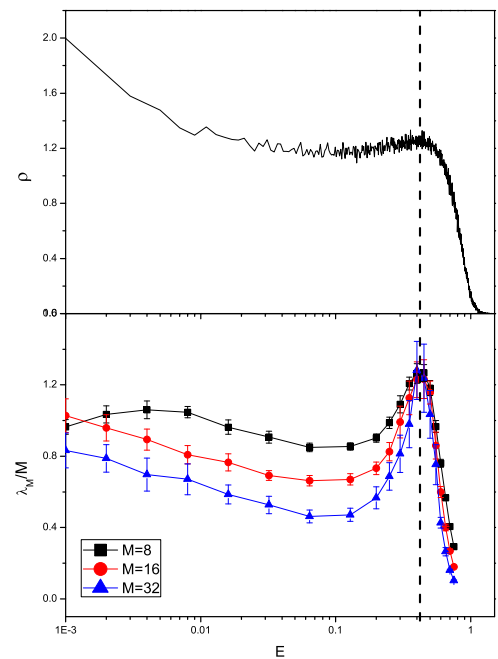}
\caption{(Color online) The DOS (top) and the localization
length in finite systems (bottom) for $g_1=g_3=0.5$, $g_0=g_2=0$.
The dashed line shows the  LD transition.} \label{g_1g_3detail}
\end{figure}

\begin{figure}
\includegraphics[scale=0.8]{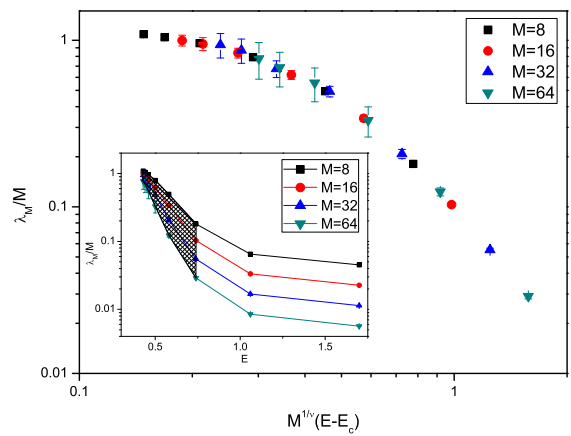}
\caption{(Color online) Scaling curve for the case $g_1=g_3=0.5$,
and $g_0=g_2=0$. Insert: the dependence of $\lambda_M/M$ on the energy $E$
for different system sizes $M$. The shaded area is used for scaling and the
critical exponent is found to be $\nu=3.23\pm 0.26$} 
\label{g_1g_3}
\end{figure}

We set $g_1=g_3=0.5$, $g_0=g_2=0$. The
density of states and the localization lengths are
plotted in Fig.~\ref{g_1g_3detail} for $E>0$; there is
symmetry under $E\to -E$. The extended states are no longer
at $E=0$ but shifted to $E=
E_c\sim\pm 0.42$. At $E_c\sim\pm 0.42$, we study the localization
properties using the data in the range of $|E|>E_c$, since data in the range
$|E|<E_c$ are close to both critical points and are likely to result in inaccurate results. The maximum system size used is
 $M=64$. Because we do not have {\em a priori} knowledge of
$E_{c}$,  the statistical procedure
discussed in Appendix~\ref{AppendixC} is employed to determine $E_c$, hence the critical
exponent $\nu$. The data collapse is shown in Fig.~\ref{g_1g_3}.
The critical exponent for this parameter set is found to be
$\nu=3.23\pm 0.26$, distinct from the nonrelativistic case of
$\nu\sim 7/3$.

The present problem  can be exactly mapped onto  the
the spin-orbit scattering involving the two-state Landau
level problem discussed in Ref.~\onlinecite{Hikami:1996}; our 
results are in full agreement. From
Fig.~\ref{g_1g_3detail}, there appears to be a divergence in the DOS at the band center, corresponding to a possible
LD transition at $E=0$. As shown above, the Cauchy distribution does lead to  a $\ln{E}$
divergence, but such a weak divergence is difficult to detect numerically; note, however, that the numerical calculation  involves Gaussian disorder. A semiclassical explanation\cite{Lee:1994}
of the existence of an extended state at the band center for the two-state Landau level problem is known. However, such an argument is delicate and   fails  if a  the third kind of disorder
is present, which is likely in graphene, where 
potential disorder can not be avoided. Thus,  we shall not consider further the
possible extended state at the band center.

\begin{figure}
\includegraphics[scale=1.2]{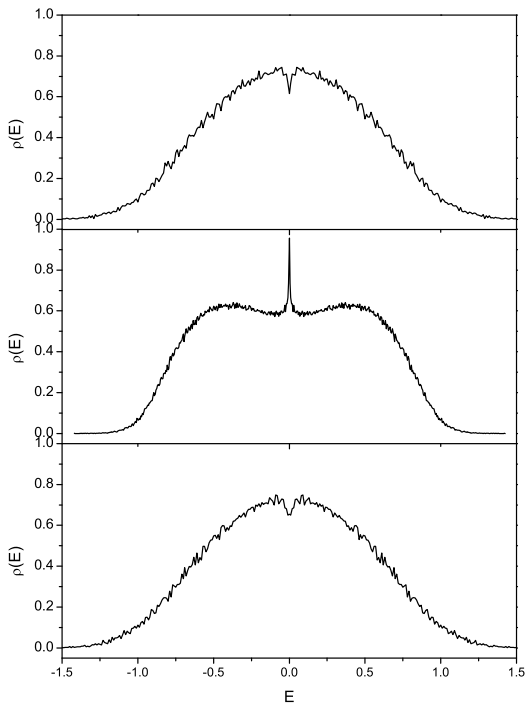}
\caption{The DOS when the mass disorder $g_3$ and
the internode coupling $g_1$ are both present. Parameters are: top:
$g_3=10g_1=0.15$; middle: $g_3=g_1=0.15$; bottom:
$g_3=0.1g_1=0.15$.} \label{dosg_1g_3}
\end{figure}

It is interesting to study the behavior as the ratio $g_{1}/g_{3}$ is varied. The result for the DOS is shown in  Fig.~\ref{dosg_1g_3}.
Note that the energy $E$ is in the unit of $2\sqrt{g_1^2+g_3^2}$. So the extent of
the band increases,  as $g_1$ increases. 
The divergence of the DOS at the band center 
is a unique feature when $g_1$ and $g_3$ are
equal, while in the 
extreme limits there may be a slight dip at $E=0$.

\begin{figure}
\includegraphics[scale=0.8]{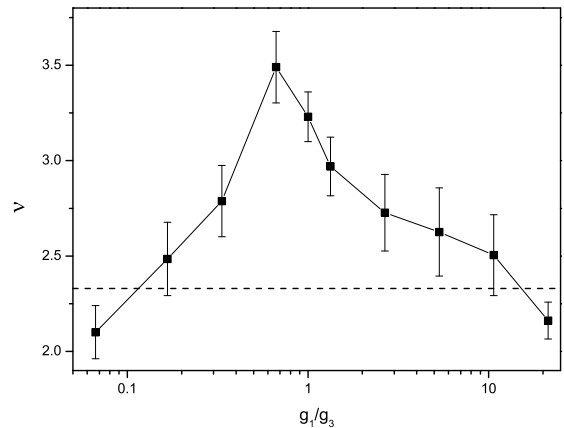}
\caption{The dependence of the  exponent $\nu$ on $g_1/g_{3}$. The
parameters are $g_0=g_2=0$, $g_3=0.15$, and $m=0$. The dashed line
corresponds to $\nu=7/3$.}
\label{g_1g_3Nu}
\end{figure}

As to  $\nu$, a continuously exponent is suggested in Fig.~\ref{g_1g_3Nu}. In the
limits $g_3\gg g_1$ or $g_3\ll g_1$, only one type of
disorder dominates, hence the value $\nu\sim 7/3$ is plausible. The
deviation from this value is the largest when $g_1\sim g_3$, although
the data collapse becomes insensitive  to the value of $\nu$ in the same
regime, resulting in larger error. Nonetheless, the results are
suggestive of a continuously varying critical exponent.

In the Hamiltonian, the mass disorder $V_3(x,y)$ and the internode scattering
disorder $V_1(x,y)$ are accompanied by the Pauli matrices $\eta_3$ and
$\eta_1$.  If we apply a unitary transformation
\begin{equation}
    T=\mathcal{C}\left(
        \begin{array}{cc}
          I & -I \\
          I & I \\
        \end{array}
      \right)
\end{equation}
corresponding to a rotation of $\pi/2$ about $y$ axis, where
$\mathcal{C}$ is a normalization factor, the disorder
Hamiltonian (\ref{hamiltonian}) will be transformed such that  $g_3\to g_1$, $g_1\to -g_3$. Because we are studying
statistical properties of the system, and all distribution functions
are symmetric about zero, the negative sign in front of the $g_3$
is of no importance. This means that this
unitary transformation effectively interchanges $g_1$ and $g_3$,
hence map the regime $g_3>g_1$ to the regime $g_3<g_1$. Note  the symmetry between the two regimes in  Fig.~\ref{dosg_1g_3}  and Fig.~\ref{g_1g_3Nu}.

\subsection{Three disorder fields}

\begin{figure}[h]
\includegraphics[scale=0.8]{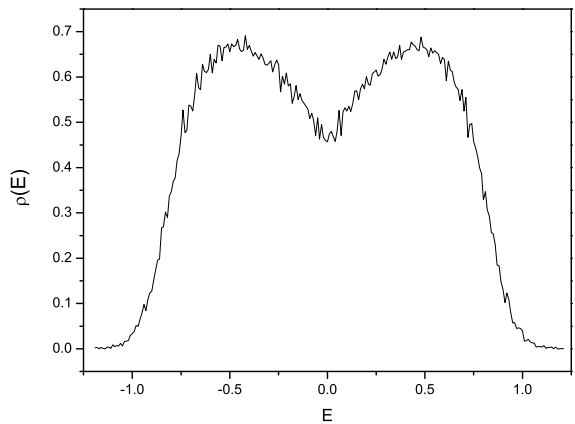}
\caption{The DOS as a function of energy $E$ in
the case $g_1=g_2=g_3=0.5$, $g_0=0$.} \label{dosg_1g_2g_3}
\end{figure}

\begin{figure}
\includegraphics[scale=0.8]{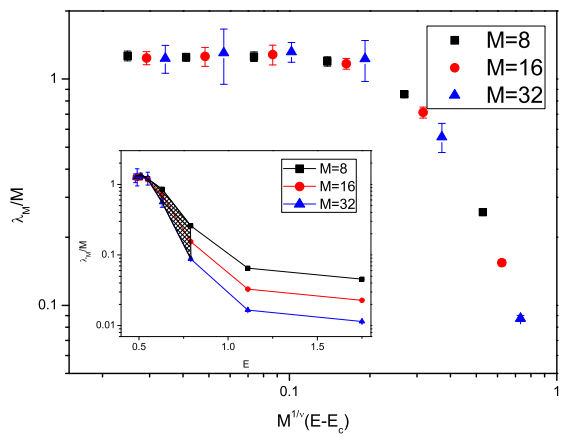}
\caption{(Color online) The scaling curve for the case $g_0=0$, and
$g_1=g_2=g_3=0.5$. Insert: the dependence of $\lambda_M/M$ on the energy $E$
for different system sizes $M$. The shaded area is used for scaling.
The critical exponent  $\nu$ is found to be $3.6\pm 0.3$}. 
\label{g_1g_2g_3}
\end{figure}

The  important case in this category is when $g_1\sim g_2\sim g_3$;  other cases can be roughly understood in terms of the cases discussed above. For numerical
computation, we take $g_1=g_2=g_3=0.5$. The DOS
is shown in Fig.~\ref{dosg_1g_2g_3}. Compared to
the  case when only $g_1=g_3=0.5$, discussed above, the divergence of the DOS at $E=0$ is missing, but
the two peaks at $E= \pm 0.46$ survive. This is
suggestive of nonexistence of extended states at the band
center, but a LD transition at $E\sim \pm
0.46$, which is confirmed by the  scaling curve shown in Fig.~\ref{g_1g_2g_3}, and a critical exponent of
$\nu=3.6\pm 0.3$ is obtained. The error bar  is large due to a substantial  degree of disorder,  but the exponent is distinctly  different from  the value $\nu=7/3$, indicating a new universality class.

\subsection{Four disorder fields}

Finally, we have also examined the case $g_{0}\ne 0$, $g_{1}\ne 0$. $g_{2}\ne 0$, $g_{3}\ne 0$. We have found that when $g_0$ is small, the
potential disorder simply broadens the density of states, and results in 
a value of $\nu$ as though it did not
exist; on the other hand a large value of $g_0$ drives $\nu$ to a value close to $7/3$.

\section{LD transition  for the massive case}
The constant mass term results in new physics when
combined with the internode coupling, and the resulting phenomena are
different from the case when the mass disorder
and the internode coupling are combined,  as in the previous section.

\subsection{One disorder field}
Consider first the case $m\ne 0$, and either $g_0$ or $g_3$ is nonzero.
From Eq.~(\ref{hamiltonian}), the random Hamiltonian matrix are block
diagonal, hence the two types of nodal fermions are uncoupled.
The energy of the two fermions are
shifted by the amount of $\pm m$, and the DOS
 is simply a superposition of two bell shaped functions
centered at $\pm m$. As to localization properties, because the
two fermions will have different critical energy, namely $E_c=\pm
m$, at a particular energy $E$ these localization lengths will be
different from each other.  The data collapse for the
fermion with $E_c=m$ once again gives a critical exponent of
$\nu\sim 7/3$.

\begin{figure}
\includegraphics[scale=0.8]{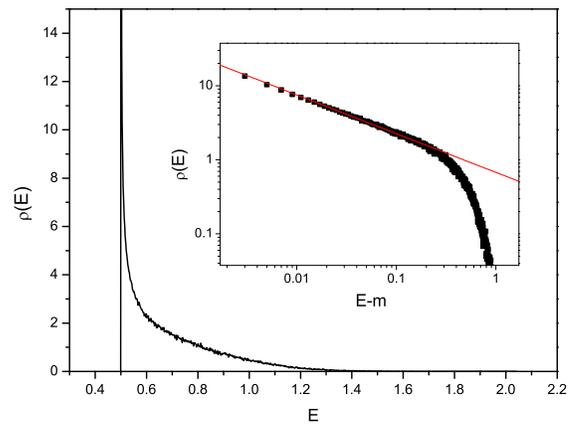}
\caption{The DOS  as a function of energy $E$. In
the case $g_0=g_2=g_3=0$, $g_1=0.5$, and $m=0.5$, the DOS vanishes
when $|E|<m$. Insert: the logarithmic plot of the DOS around $E=m$,
and $m=0.5$. The best fit gives a slope of $-0.523\pm.003$.}
\label{dosg_1m}
\end{figure}

Consider now finite mass $m\ne 0$ and only one
internode coupling, for example $g_1\ne 0$. The calculated DOS
with $E>0$ region is shown in Fig.~\ref{dosg_1m} for the parameter
set $g_0=g_2=g_3=0$, $g_1=0.5$, and $m=0.5$. Qualitatively the results are
similar to the analytical calculations involving the Cauchy distribution, even though the numerical
computation is for the Gaussian distribution of disorder.  First, the DOS vanishes in the
region $|E|<m$. Second,  the
analytical calculation shows that
$\rho(E)\sim(E-m)^{-1/2}$, $E\to m^+$, independent of the value of
$g_1$. The  insert of Fig.~\ref{dosg_1m} yields an
exponent of $-0.523\pm.003$. We have checked that this value
does not vary with $g_1$, within our numerical accuracy.

\begin{figure}
\includegraphics[scale=0.8]{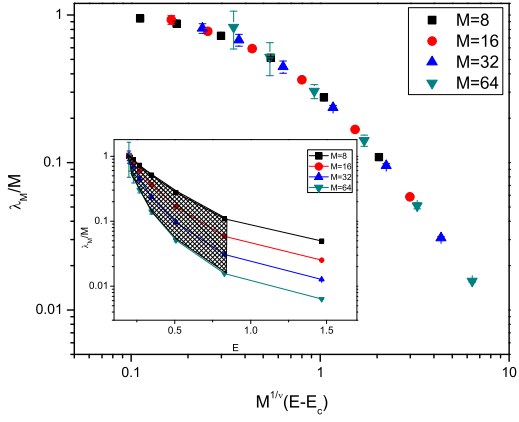}
\caption{(Color online) The scaling curve for the case $g_0=g_2=g_3=0$,
$g_1=0.4$, and $m=0.15$. Insert: dependence of $\lambda_M/M$ on
energy $E$ for different system sizes $M$. The shaded area is used for
scaling. Critical exponent is found to be $\nu=1.82\pm 0.06$, with
the choice of $E_c=m$. } \label{g_1m}
\end{figure}

\begin{figure}
\includegraphics[scale=0.8]{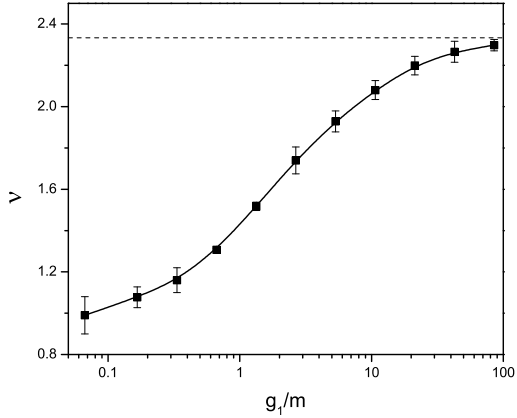}
\caption{The dependence of the  critical exponent $\nu$ on
 $g_1$ (normalized by
the mass $m$). The parameters are $g_0=g_2=g_3=0$, and $m=0.15$. The dashed
line indicates the level of $\nu=2.33$.} \label{g_1mNu}
\end{figure}

As to LD transition, we perform data collapse
with  $E_c=m$. The critical exponent turns out to be
$\nu=1.82\pm 0.06$ for the parameters $g_0=g_2=g_3=0$, $g_1=0.4$,
and $m=0.15$, see Fig.~\ref{g_1m}, which is significantly different from the usual case
corresponding to $\nu\approx 7/3$. This striking result implies that
the system belongs to a new universality class. We now vary $g_1$, keeping $m$
fixed to $0.15$, and the result for the exponent $\nu$ is shown in Fig.~\ref{g_1mNu}. 
It appears that the exponents continuously vary with the ratio $g_{1}/m$.

\subsection{Two disorder fields}
\begin{figure}
\includegraphics[scale=1.5]{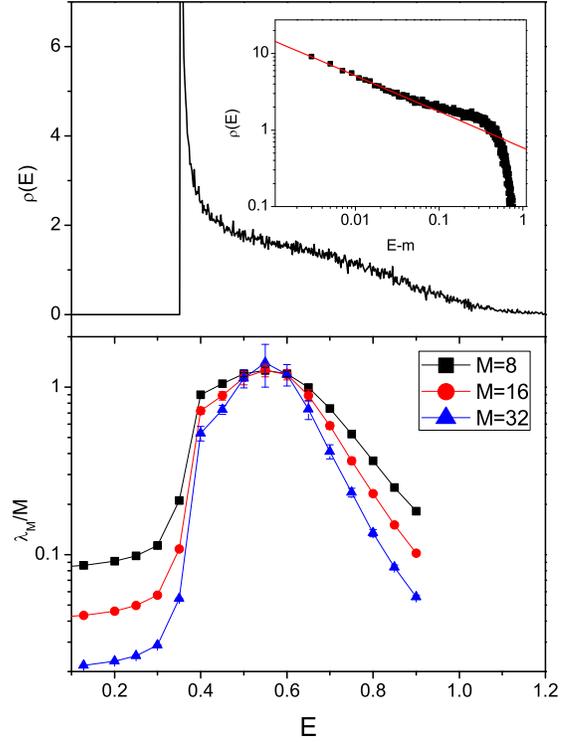}
\caption{(Color online) The DOS (top) and the localization
length in finite systems (bottom) for $g_1=g_2=m=0.5$, $g_0=g_3=0$.
Insert in the top: the logarithmic plot of the DOS around $E=m$. The
best fit gives a slope of $-0.47\pm.01$.} \label{g_1g_2mdetail}
\end{figure}

The most relevant case
corresponding to graphene is the one with two types of
internode scattering of comparable magnitude. Therefore, we choose 
$g_1=g_2=m=0.5$. The DOS is shown
in Fig.~\ref{g_1g_2mdetail}. Note that the energy is now  measured in
units of $2\sqrt{g_1^2+g_2^2}=\sqrt{2}$, so that the divergence
is  located at $E=m$, which is $m=0.5/\sqrt{2}=0.354$. From the insert in
Fig.~\ref{g_1g_2mdetail}, we find  a slope of
$-0.47\pm.01$, which is consistent with the 
analytical result  for the Cauchy distribution, namely,
$\rho(E)\sim\ln|E-m|/\sqrt{|E-m|}$. Also note the gap in the DOS.

\begin{figure}
\includegraphics[scale=0.8]{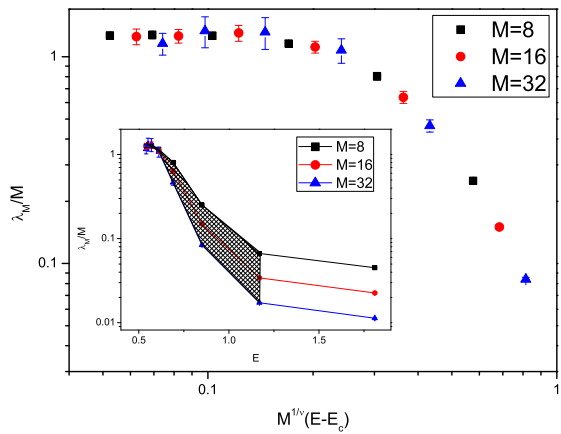}
\caption{(Color online) The scaling curve for the case $g_1=g_2=m=0.5$,
$g_0=g_3=0$. Insert: the dependence of $\lambda_M/M$ on the energy $E$ for
different system sizes $M$. The shaded area is used for scaling.}
\label{g_1g_2m}
\end{figure}

The crossing point in Fig.~\ref{g_1g_2mdetail} indicates 
 $E_c\sim 0.55$ instead of $E_c=m$. The data collapse is
shown in Fig.~\ref{g_1g_2m} with a critical exponent of $\nu=3.8\pm
0.2$. This critical exponent is not close to any of the values found for
a finite mass with a single internode coupling, as in
Fig.~\ref{g_1mNu}. However, it is reasonably close to the exponent for $m=g_0=g_2=0$, and $g_1\sim g_3$  (see
Fig.~\ref{g_1g_3Nu}), which is also equivalent to the case
$m=g_0=g_3=0$, and $g_1\sim g_2$, as discussed above. Note that
$m=0.5$ is much smaller than the bandwidth
$2\sqrt{g_1^2+g_2^2}=1.414$, and this critical exponent indicates that
the presence of small  finite mass will have little
effect on the critical exponent as long as two internode couplings
are finite.

At the other limit, when  $m$ is large enough compared to
the band width,  the critical exponent $\nu\to
1$. Thus, it is reasonable to believe that the exponent also
varies continuously, as a function of $g_1/m$, provided that $g_1=g_2$,
and the behavior is similar to in Fig.~\ref{g_1m}, except that
$\nu\to 3.8$ in the limit $m\to 0$.

\subsection{Three disorder fields}

Because potential
disorder is always present  in experiments on graphene, we would like to discuss the case when $g_0\neq
0$, $m\neq 0$, and $g_1=g_2\neq 0$. 

First, when $m$ is the smallest parameter, it
can be neglected, and hence the massless
case discussed above is recovered. Our
numerical computations gave a critical exponent of $\nu \sim 3.8$
when $m \ll g_0 \ll g_1=g_2$, and $\nu\sim 2.3$ in the limit
$m \ll g_1=g_2\ll g_0$, because now the potential disorder is
more important than the rest.

When $g_0$ is the smallest, it
will have little effect. Therefore, as discussed
in the previous subsection, there will be a continuously varying exponent
from $\nu\sim 1.0$ for $g_0\ll g_1=g_2 \ll m$ to $\nu\sim 3.8$ for
$g_0\ll m \ll g_1=g_2$.

Finally, if $g_1=g_2$ are smaller than the rest, the internode scattering
 is no longer important, and hence the two nodal fermions
will be decoupled. The exponent will therefore be always
$\nu\sim 2.3$ regardless of the relationship between $m$ and $g_0$.

\section{Conclusions}
We have analyzed the effects of
disorder  on the LD transition in the LLL of graphene. Because both 
types of  internode scattering, present in the LLL, arise from the
random hopping, they will have roughly the same strength. Because
the sources of the mass disorder and the internode
scattering are different, their strengths will be generically different.  In some special cases of disorder
combinations we have found new universality classes of LD
transition in contrast to the conventional IQH.

Our results for the LD transitions in the LLL have direct
experimental relevance for the plateau transitions in graphene. Consider first
 the cases where both the spin and the nodal degeneracies
are completely removed. A number of authors have shown
 that the inclusion of a finite mass and Zeeman energy can
explain the appearance of plateaus at $\nu_{f}=0,\pm1, \pm 2q$.

Because experiments resolve the spin and nodal
 splitting, intranode scattering which always broadens the Landau
 levels is weak compared to $E_{z}$ and $m$.  If the internode scattering
 strength is larger than the intranode scattering strength,
 we expect that in the lowest Landau level $0\rightarrow\pm1$, $1\rightarrow 2$ and $-1\rightarrow -2$ plateau
 transitions can have different universality classes, in contrast to the conventional
 IQH effect. In the opposite limit, when the intranode scattering is considerably stronger,
 the plateau transitions will fall into the conventional IQH universality class with $\nu \sim 2.3$.

For the spin and nodally degenerate plateaus, the potential scattering is strong compared
to the Zeeman energy and the mass gap. Theoretically, from our analysis of the massless cases we can infer
the plateau transitions to be of the conventional IQH type. However, in experiments, if scaling
with respect to the band center is invoked, an effective exponent for these plateau transitions will be observed.

Plateaus at $\nu_{\textrm{f}}=\pm 4, \pm6,..$ involve higher Landau levels. In the higher Landau levels
both the random potential at the lattice scale and the random hopping will have nonzero intranode
as well as internode scattering contributions. This will complicate the analysis of LD
transitions in these levels. Because inclusion of a finite mass does not lift the nodal
degeneracies of higher Landau levels, the effect of  internode scattering
can be strong. In view of the degeneracy factor, there is the possibility of observing an
effective exponent for these higher plateau transitions.

\vspace{5cm}

\begin{acknowledgments}
This work was  was supported by the NSF under the Grant No. DMR-0411931. 
\end{acknowledgments}

\appendix
\section{Integral}
\label{AppendixA}
For the Cauchy distribution, in the absence of mass disorder, the
calculation of the DOS in
Eq.~(\ref{eq:massdencauch}) involves the integral
\begin{widetext}
\begin{equation}
I=\frac{1}{4}\int_{0}^{\infty} dx \int_{0}^{\infty} dy \int_{0}^{2
\pi} d\alpha_{1} \int_{0}^{2\pi} d\alpha_{2} \exp
\bigg[-ax-by-2g_{2}\sqrt{xy}|\sin(\alpha_{1}-\alpha_{2})|-2g_{1}\sqrt{xy}|\cos(\alpha_{1}-\alpha_{2})|\bigg],
\end{equation}
\end{widetext}
where $a=g_{0}-i\epsilon_{1}$ and $b=g_{0}-i\epsilon_{2}$. In the
massless case $a=b=g_{0}-i\epsilon$. After performing the
integrals over one of the angles,the double integral over the
angles is reduced to
\begin{equation}
I_{ang}=8\pi \int_{0}^{\pi /2}d\alpha_{1} \exp[-2\Delta
\sqrt{xy}\cos(\alpha_{1}-\beta)],
\end{equation}
where $\Delta=\sqrt{g_{1}^{2}+g_{2}^{2}}$ and $\tan
\beta=g_{2}/g_{1}$. Now, expanding the exponential in a power
series, integrals over $x$ and $y$ can be easily performed. For the
angular integral we use the relation
\begin{widetext}
\begin{equation}
\int_{0}^{\pi /2}
d\alpha_{1}\cos^{l}(\alpha_{1}-\beta)=-\frac{1}{l+1}\bigg[\sin^{l+1}
\beta {}_{2}F_{1}(\frac{l+1}{2},\frac{1}{2},\frac{3+l}{2},\sin^2
\beta)+\cos^{l+1}
\beta{}_{2}F_{1}(\frac{l+1}{2},\frac{1}{2},\frac{3+l}{2},\cos^2
\beta)\bigg],
\end{equation}
\end{widetext}
where ${}_{2}F_{1}$ is Gauss's hypergeometric function and obtain
\begin{eqnarray}
I&=&-\frac{2 \pi}{\Delta ab}
\sum_{j=1}^{2}\sum_{l=0}^{\infty}\frac{\Gamma^{2}((l/2)+1)}{\Gamma(l+2)}g_{j}(\frac{-2g_{j}}{\sqrt{ab}})^l
\nonumber \\
& & \times
{}_{2}F_{1}(\frac{l+1}{2},\frac{1}{2},\frac{3+l}{2},g_{j}^{2}/\Delta^{2}).
\end{eqnarray}
The summation over $l$ can
be performed by the following trick: (1) use the integral
representation of ${}_{2}F_{1}$, (2) perform a power series
summation which is simple in these cases, and (3) complete the
integration over the auxiliary variable introduced for the
integral representation. For simplicity we will specialize to the 
cases: (i) $g_{1}\neq0, g_{2}= 0$ and (ii) $g_{1}=g_{2}=g_{IN}$.

(i) $g_{1}\neq0, g_{2} = 0$: In this case we have
\begin{eqnarray}
 I&=&-\frac{2 \pi}{
ab}\sum_{l=0}^{\infty}\frac{\Gamma^{2}((l/2)+1)}{\Gamma(l+2)}(\frac{-2g_{1}}{\sqrt{ab}})^l
\nonumber \\
& & \times {}_{2}F_{1}(\frac{l+1}{2},\frac{1}{2},\frac{3+l}{2},1)
\end{eqnarray}
We now use the following two relations
\begin{equation}
{}_{2}F_{1}(\frac{l+1}{2},\frac{1}{2},\frac{3+l}{2},1)=\frac{\sqrt{\pi}\Gamma((3+l)/2)}{\Gamma((l/2)+1)},
\end{equation}
\begin{equation}
\sum_{l=0}^{\infty}(-x)^{l}\frac{\Gamma((l/2)+1)\Gamma((3+l)/2)}{\Gamma(l+2)}=\frac{\sqrt{\pi}}{2+x},
\label{eq:sum}
\end{equation}
to obtain Eq.~(\ref{eq:int1}).

(ii) $g_{1}=g_{2}=g_{IN}$: In this case we have
\begin{eqnarray}
 I&=&-\frac{2 \sqrt{2}\pi}{
ab}\sum_{l=0}^{\infty}\frac{\Gamma^{2}((l/2)+1)}{\Gamma(l+2)}(\frac{-2g_{IN}}{\sqrt{ab}})^l
\nonumber \\
& & \times
{}_{2}F_{1}(\frac{l+1}{2},\frac{1}{2},\frac{3+l}{2},1/2).
\end{eqnarray}
Using the integral representation
\begin{eqnarray}
{}_{2}F_{1}(\frac{l+1}{2},\frac{1}{2},\frac{3+l}{2},\frac{1}{2})&=&\frac{\Gamma((3+l)/2)}{\Gamma(1/2)\Gamma((l/2)+1)}\int_{0}^{1}
dtt^{-1/2}
\nonumber \\
& &\times (\frac{1-t}{1-t/2})^{l/2}(1-\frac{t}{2})^{-1/2},\nonumber \\
\end{eqnarray}
and Eq.~(\ref{eq:sum})we get
\begin{equation}
I=-{\pi \sqrt{2}}{\sqrt{ab}}\int_{0}^{1}
\frac{dt}{\sqrt{ab}\sqrt{t(1-(t/2))}+g_{IN}\sqrt{t(1-t)}}
\end{equation}
After performing the integral over $t$ we get Eq.~(\ref{eq:int2}).
After setting $m=0$ one recovers Eq.~(\ref{eq:integral1})

\section{Recursive Green's Function}
\label{AppendixB}

Because there are  two types of fermions in our problem, the
recursive Green's function technique is a little more complicated than
that introduced by Huckenstein\cite{Huckestein:1995}.  All the
matrix elements, such as those in Eqn.~(6.9) in
Ref.~\onlinecite{Huckestein:1995}, become $2\times 2$ matrices;
hence, all the operations, such as the multiplication and inversion  are matrix operations.

Denote the $2\times 2$ matrix $\langle i|G|j\rangle$ simply by
$G(i,j)$. Suppose $G^{(K)}(i,j)$, $i,j=1,\ldots,K$, the
Green's function containing $K$ momentum states is
available, then as we add another momentum state, the recursion
relations for $G^{(K+1)}(i,j)$ are:
\begin{widetext}
\begin{equation}\label{recursion}
\begin{split}
    G^{(K+1)}(K+1,K+1)&=\left(E-V(K+1,K+1)-\sum_{i,j} V(i,K+1)^\dag
    G^{(K)}(i,j)V(j,K+1)\right)^{-1}\\
    G^{(K+1)}(i,K+1)&=\left[\sum_j
    G^{(K)}(i,j)V(j,K+1)\right]G^{(K+1)}(K+1,K+1)\mbox{\quad:\quad}i\le
    K\\
    G^{(K+1)}(i,j)=G^{(K)}(i,j)&+G^{(K+1)}(i,K+1)G^{(K+1)}(K+1,K+1)^{-1}G^{(K+1)}(K+1,j)\mbox{\quad:\quad}i,j\le
    K
\end{split}
\end{equation}

These matrix inversions here can be accurately computed because the
 sizes are small. Corresponding to the two types of fermions, we
are interested in  $G^{(K)}(1,K)_{n,n}$,
$n=1,2$, because the localization length of a system of width $M$
for the $n^{th}$ fermion, $\lambda_{M,n}$, is related to this
quantity by,
\begin{equation}
\begin{split}
    \lambda_{M,n}^{-1}&=-\frac{M}{K\sqrt{2\pi}l_B}\ln|G^{(K)}(1,K)_{n,n}|\\
    &=-\frac{M}{K\sqrt{2\pi}l_B}\sum_{k=1}^K\ln|q^{(K)}_n| ,
\end{split}
\end{equation}
where $q^{(K)}_n=G^{(K)}(1,K)_{n,n}/G^{(K-1)}(1,K-1)_{n,n}$.
Furthermore, if we define a set of $2\times 2$ matrices
$g^{(K)}(j)$, $j=1,\ldots,K$, such that their elements
$g^{(K)}(j)_{m,n}=G^{(K)}(1,j)_{m,n}/G^{(K)}(1,K)_{m,m}$, from the
recursion relations (\ref{recursion}), we obtain:

\begin{equation}
\begin{split}
    q^{(K+1)}_n&=\left[\left(\sum_{j}g^{(K)}(j)V(j,K+1)\right)G^{(K+1)}(K+1,K+1)\right]_{n,n}\\
    g^{(K+1)}(i)_{m,n}&=\frac{1}{q^{(K+1)}_m}\left[g^{(K)}(i)_{m,n}+\left(\sum_j
    g^{(K)}(j)V(j,K+1)G^{(K+1)}(K+1,i)\right)_{m,n}\right]\mbox{\quad:\quad}i\le K\\
    g^{(K+1)}(i)_{m,n}&=\frac{1}{q^{(K+1)}_m}\left(\sum_j
    g^{(K)}(j)V(j,K+1)G^{(K+1)}(K+1,i)\right)_{m,n}\mbox{\quad:\quad}i=K+1\\
\end{split}
\end{equation}
\end{widetext}

\section{Data collapse}
\label{AppendixC}
In this appendix, we describe how we can extract the
the exponent $\nu$, and the critical energy $E_c$,
if necessary, based on the computed localization lengths in finite
systems $\lambda_M(E)$ assuming a single parameter scaling assumption.

Suppose we have obtained $\{\lambda_M(E)/M\}$ in systems with
$\{M_i\}_{i=1}^{N_M}$ for $\{E_j\}_{j=1}^{N_E}$, each with a
standard deviation $\{\sigma_{M_i,E_j}\}$. Our goal is to find
out the proper values of $\nu$ and $E_c$ such that all the
$N_E\times N_M$ data points collapse on to a single curve:
\begin{equation}
    \frac{\lambda_M(E)}{M}=f\left(M^{1/\nu}(E-E_c)\right)
\end{equation}
Since $f(x)$ is unknown, it is difficult to characterize the quality
of the data collapse. To overcome this difficulty, we proceed as follows: suppose that we are
given a pair of values $(\nu,E_c)$, we can attempt to represent the
unknown function $f(x;\nu,E_c)$ by a polynomial of degree $N$ by
simply performing a general fit to Eq.~(\ref{fitlambda}) given
below, based on a total of $N_E\times N_M$ data points
$\{(x,y,\sigma_y):(\log[M_j^{1/\nu}(E_i-E_c)],\log[\lambda_M(E_i)/M_j],\sigma_{M_j,E_i})\}$:
\begin{equation}
    \log\frac{\lambda_M(E)}{M}=\sum_{k=0}^N a_k \left[\log(M^{1/\nu}(E-E_c)) \right ]^k ,
    \label{fitlambda}
\end{equation}
where $\{a_i\}_{i=0}^N$ are the coefficients to be fitted. In the
computer implementation, the order of polynomials was chosen to be
$N=5$, since no significant changes were noted by increasing $N$ to
9. The quality of this fit~\cite{Press:1992} is represented by the
variable $S$ defined as:
\begin{equation}
    S(\nu,E_c)=\sum_{i=1}^{N_E\times
    N_M}\left[\frac{y_i-f(x_i;\nu,E_c)}{\sigma_i}\right]^2
\end{equation}
If the preset values $(\nu,E_c)$ are not correct, the data points
will be scattered, resulting in a large value of $S$, which in turn
indicates a poor data collapse. However, when $(\nu,E_c)$ attain the
correct localization length exponent and the correct critical energy,
respectively, $S$ will be minimized. Following this procedure, by
minimizing $S$ with the standard gradient descent method, we are
able to determine correctly both the critical energy $E_c$ and the
localization length exponent $\nu$. Because the 
scaling law is only valid in the close vicinity of the critical energy, once
$E_c$ is obtained from above procedure, we have to check, for the
purpose of self consistency, that all energies used in the data
collapse are indeed close to $E_c$.

As for the statistical error of $\nu$, the usual procedure is to
assume that the minimized $S_{min}$ follows a $\chi^2$ distribution,
and hence the error bar can be drawn corresponding to a certain
confidence probability. However, this is not the case in this
problem, since $S_{min}$ does not follow the $\chi^2$ distribution
due to the nonlinear form of the estimated parameters $\nu$ and $E_c$ in
(\ref{fitlambda}).\cite{Press:1992} To draw an error bar for $\nu$
statistically correctly, recall that we have the original data
$\{(x_i,y_i,\sigma_{y_i})\}$. We generate a large number of data
sets synthetically $\{(x_i^{(k)},y_i^{(k)},\sigma_{y_i}^{(k)})\}$,
for $k=1,2,\ldots, N_s$, such that $x_i^{(k)}=x_i$,
$\sigma_{y_i}^{(k)}=\sigma_{y_i}$, and $y_i^{(k)}$ a variable
randomly distributed in the Gaussian form with a mean of $y_i$ and a
standard deviation of $\sigma_{y_i}$. Next, we perform exactly the
same procedure to get $\nu^{(k)}$ for each synthetic data set
$\{(x_i^{(k)},y_i^{(k)},\sigma_{y_i}^{(k)})\}$, as was performed in
actual data set $\{(x_i,y_i,\sigma_{y_i})\}$ for estimating the
$\nu$ and $E_c$. Finally, the error bar for $\nu$ is drawn as the
estimated standard deviation:
\begin{equation}
    \sigma_{\nu}=\left[\frac{1}{N_s-1}\sum_{k=1}^{N_s}(\nu^{(k)}-\overline{\nu^{(k)}})^2\right]^{1/2}
\end{equation}
where $N_s$ is the number of synthetic data sets, and
$\overline{\nu^{(k)}}$ is the average of $\nu^{(k)}$. $N_s=10^4$ in
computer implementation.

%\bibliography{reference}

\end{document}